\documentclass[
reprint,
superscriptaddress,
amsmath,amssymb,
aps
]{revtex4-2}

\usepackage{graphicx}
\usepackage{dcolumn}
\usepackage{bm}
\usepackage{hyperref}
\usepackage[version=4]{mhchem}    
\usepackage{csquotes}
\hypersetup{colorlinks=true,linkcolor=red,citecolor=blue}
\usepackage{float}
\usepackage{physics}
\usepackage{xspace}
\usepackage[normalem]{ulem}

\usepackage{lipsum}

\usepackage{prettyref}
\newcommand{\pref}[1]{\prettyref{#1}}
\newrefformat{fig}{Fig.~\ref{#1}}
\newrefformat{tab}{Table~\ref{#1}}
\newrefformat{sec}{Sec.~\ref{#1}}
\newrefformat{ssec}{Sec.~\ref{#1}}
\newrefformat{app}{App.~\ref{#1}}
\newrefformat{eqn}{Eq.~(\ref{#1})}

\usepackage{bbding} 

\usepackage[dvipsnames,table]{xcolor}

\newcommand{\im}{\mathrm{i}}

\newcommand{\editor}[2]{%
  \expandafter\newcommand\csname #1note\endcsname[1]{%
    \textcolor{#2}{(\textbf{#1:} ##1)}}%
  \expandafter\newcommand\csname #1\endcsname[1]{%
    \textcolor{#2}{##1}}%
  \expandafter\newcommand\csname #1cancel\endcsname[1]{%
    \textcolor{#2}{\sout{##1}}}%
  \expandafter\newcommand\csname #1change\endcsname[2]{%
    \textcolor{#2}{\sout{##1} ##2}}%
  \newenvironment{#1text}{\color{#2}}{\color{black}}
}

\editor{TI}{magenta}

\editor{R}{red}

\begin{document}

\title{Strong correlations and local self-energies from on-site ensembles}

\author{Alberto Carta} \email{alberto.carta@psi.ch}
\affiliation{PSI Center for Scientific Computing, Theory, and Data, Paul Scherrer Institute, 5232 Villigen PSI, Switzerland}
\author{Hugo U.~R.~Strand} 
\affiliation{School of Science and Technology, \"Orebro University, SE-70182 \"Orebro, Sweden}
\author{Michael Sch\"uler}
\affiliation{PSI Center for Scientific Computing, Theory, and Data, Paul Scherrer Institute, 5232 Villigen PSI, Switzerland}
\affiliation{Department of Physics, University of Fribourg, 1700 Fribourg, Switzerland}
\author{Nicola Marzari} 
\affiliation{PSI Center for Scientific Computing, Theory, and Data, Paul Scherrer Institute, 5232 Villigen PSI, Switzerland}
\affiliation{Theory and Simulation of Materials (THEOS), and National Centre for Computational Design and Discovery of Novel Materials (MARVEL), École Polytechnique Fédérale de Lausanne, 1015 Lausanne, Switzerland}
\affiliation{U Bremen Excellence Chair, Bremen Center for Computational Materials Science, and MAPEX Center for Materials and Processes, Universität Bremen, 28359 Bremen, Germany}
\affiliation{Theory of Condensed Matter, Cavendish Laboratory, University of Cambridge, Cambridge, CB3 0US, United Kingdom}

\date{\today}

\begin{abstract}
Addressing the many-body electronic-structure problem is a central goal of modern condensed‑matter physics. Paramagnetic Mott insulators, in particular, have long  represented a challenge for standard approaches, such as density-functional theory. Historically, these systems have been tackled either by considering many‑body dynamics, as in the case of dynamical mean-field theory (DMFT), or, more recently, by invoking a polymorphous description consisting of large supercells populated with static symmetry‑broken motifs whose spatial average restores the paramagnetic state.
Inspired by these viewpoints, we introduce the on‑site dephased ensemble (DE) approximation, in which the local electronic-structure problem is described by a thermal ensemble of all accessible local static solutions; this gives rise to a strong frequency dependence of the local electronic self-energy, as seen in DMFT or in the coherent-potential approximation of disordered alloys. We show that the DE successfully recovers defining hallmarks of strongly correlated Mott systems,  both in terms of the self‑energy as well as the persistence of local moments in time, in quantitative agreement with DMFT. By bypassing expensive quantum Monte Carlo solvers or large supercell calculations, this approach offers efficient routes to treating paramagnetic Mott systems in a first-principles setting, and highlights a deeper connection between the physics of strong correlations and that of disorder.
\end{abstract}


\maketitle

Describing the behavior of interacting electrons from first principles is one of the ultimate goals of modern electronic-structure theory; this is particularly challenging in materials with partially filled $d$ and $f$ shells. 
In particular, capturing the paramagnetic Mott insulating state, where a gap emerges from electron–electron interactions in the absence of magnetic order, has historically presented a challenge.
A straightforward application of Kohn-Sham density-functional theory (DFT) within local or semi-local energy functional approximations~\cite{Kohn:1965, Hohenberg:1964, perdew_generalized_1996} notoriously mislabels many of these systems, such as the paradigmatic transition-metal (TM) monoxides (NiO, FeO, etc.) as metals ~\cite{trimarchi_polymorphous_2018a, mazin_insulating_1997, zunger_symmetry_2026}.
To recover the experimentally observed ~\cite{bowen_electrical_1975, roth_magnetic_1958} insulating gap, it is common to impose an artificial static (e.g., antiferromagnetic) order, often within the DFT+$U$ framework, to explicitly break symmetry while correcting self-interaction errors~\cite{anisimov_band_1991a,anisimov_bandstructure_1990, mazin_insulating_1997}. While phenomenologically successful for predicting various observables~\cite{timrov_accurate_2022b, timrov_unraveling_2023}, this proxy sidesteps the true nature of the macroscopic paramagnetic phase.

One avenue to describe paramagnetic Mott systems is to combine DFT with dynamical mean-field theory (DMFT)~\cite{georgesDynamicalMeanFieldTheory1996, kotliar_electronic_2006, held_electronic_2007a}. In DMFT, the atomic many-body problem is described in terms of temporal fluctuations of local moments; these fluctuations manifest themselves as a strong frequency dependence in the local electronic self-energy $\Sigma(\omega)$, which is considered a hallmark of “strong correlations”.
A recent alternative approach ~\cite{abrikosov_recent_2016, alling_effect_2010, trimarchi_polymorphous_2018a, zunger_symmetry_2026, malyi_rise_2023a, varignon_mott_2019} exploits large supercells that host a disordered arrangement of local symmetry‑broken motifs (such as bond disproportionation or local magnetic moments), whose spatial average restores the macroscopic paramagnetic state. Combined with self‑interaction‑corrected functionals like DFT+$U$ or R2SCAN~\cite{cococcioni_linear_2005, furness_accurate_2020}, this ``polymorphous" description naturally yields an insulating state without explicit time‑dependent fluctuations. This viewpoint questions the necessity of invoking strong correlations and argues that the erroneous metallic prediction stems from artificially restricting the material to a single ``monomorphous" (i.e. symmetry‑unbroken) configuration.
The many‑body picture of temporal fluctuations and the polymorphous picture of spatial heterogeneity are each remarkably successful. Yet, whether these two viewpoints share a common theoretical ground remains an unresolved question~\cite{zunger_symmetry_2026, varignon_mott_2019, trimarchi_polymorphous_2018a, rivera_identifying_2026, Essl2026}. 

In this Letter, we take inspiration from both viewpoints to construct the on‑site dephased ensemble (DE). Starting from a generic embedding problem, instead of taking only one single-determinant mean-field solution, DE retains a thermodynamic mixture of all available local mean-field states. This minimal extension captures the defining fingerprints of strongly correlated paramagnetic Mott insulators: a pronounced frequency dependence of the local self‑energy and long‑lived local moments, and both are found to be in quantitative agreement with the DMFT solution. This approach sidesteps expensive impurity solvers as well as large supercells, and constitutes an enlightening step in understanding and connecting the many-body and the disordered polymorphous descriptions of realistic correlated materials.

We begin by considering a generic system where standard DFT fails to adequately capture the physics of localized states belonging to specific atoms (which we call ``correlated atoms"). These states together form the ``correlated subspace", a subset of the total electronic Hilbert space.
In quantum embedding approaches~\cite{Marzari_2025, Ferretti2024, georgesDynamicalMeanFieldTheory1996, kotliar_electronic_2006, held_electronic_2007a, Savrasov2004}, one focuses on a fragment of the periodic solid, a single or a few correlated atoms, embedded in a bath that represents the rest of the crystal.
For a single-atom fragment, the local screened Coulomb interaction can be described by a four-index tensor $W_{ijkl}$ (the subscripts denote orbital indices)\cite{altland_condensed_2010a, kotliar_electronic_2006, Savrasov2004, held_electronic_2007a}, which we parametrize here with the form of Liechtenstein~\cite{Liechtenstein:1995} in terms of the Hubbard $U$ and Hund’s $J$.
Finding the full solution of the embedding problem at a given temperature then amounts to evaluating the fragment’s partition function $\mathcal{Z}_{\rm loc}$.
The simplest approximation is the static mean-field (single-determinant) decoupling, also often called Hartree-Fock (HF) decoupling in the many-body literature~\cite{Martin2016, altland_condensed_2010a, kotliar_electronic_2006, georgesDynamicalMeanFieldTheory1996, carta_explicit_2025}.
In this approximation, the local self-energy takes the static form $\bar{\Sigma} = \mathcal{U}\bar{\rho}$, 
where $\bar{\rho}$ is the atomic density matrix (we use the barred notation to indicate static quantities) and $\mathcal{U}$ is defined as $\mathcal{U}_{ijkl}=W_{ijkl}-W_{ilkj}$. The HF form of the self-energy is equivalent to approximating the partition function as $\mathcal{Z}_\text{loc} \sim \exp(-\beta\Omega[\bar\Sigma])$ where the embedding grand potential $\Omega[\bar\Sigma]$ of the solution reads:
\begin{equation}
\Omega[\bar{\Sigma}] = - \frac{1}{\beta} \mathrm{Tr} \ln \bigl[ G_0^{-1} - \bar{\Sigma} \bigr] - \frac{1}{2} \mathrm{Tr} \, \bar{\rho}\,\mathcal{U}\,\bar{\rho} .
\label{eqn:grand_pot_hf}
\end{equation}
Here, $G_0$ is the non-interacting part of the embedding problem, $\beta$ is the inverse temperature and $\bar\Sigma$ constitutes a stationary point of the embedding grand potential $\delta\Omega/\delta\bar\Sigma=0$.
While simple, a single Slater determinant per correlated atom could be a very severe approximation. The static mean‑field decoupling has a well-documented tendency to lower the energy by breaking symmetry and polarizing spin and orbital densities~\cite{Martin2016, Coulson1949, Nesbet2004}; consequently, realistic multi‑orbital systems can exhibit a highly non‑convex energy landscape containing hundreds of symmetry‑broken stationary solution,  as has been shown in~\cite{ponet_energy_2024, haddadi_exploring_2026}. Static treatments, such as for instance DFT+U, select only one of these minima depending on the initialization~\cite{ponet_energy_2024, Dorado2013, Ylvisaker2009}. This locks the system into a specific symmetry‑broken basin and discards the thermodynamic weight of the many other low‑lying, often degenerate, configurations.

The central approximation we introduce in this work, which we call on-site dephased ensemble, consists in restoring the thermodynamic weight of these discarded states: we approximate the partition function of the local problem by summing over all accessible stationary points of each correlated atom
\begin{equation}
\mathcal{Z}_{\text{loc}} \approx \sum_{s} e^{-\beta \Omega[\bar{\Sigma}^{s}]}, \qquad
\left.\frac{\delta \Omega}{\delta \bar{\Sigma}}\right|_{\bar{\Sigma}^s} = 0,
\label{eqn:de_definition}
\end{equation}
where the index $s$ labels each different stationary solution with its respective self-energy $\bar\Sigma^s$.
In this approximation, the local Green's function $G_\text{DE}$ takes the compelling form:
\begin{equation}
G_{\text{DE}}(\mathrm{i}\omega_n) = \frac{1}{\mathcal{Z}_{\text{loc}}} \sum_{s} G^{s}(\mathrm{i}\omega_n) \, e^{-\beta \Omega[\bar{\Sigma}^s]},
\label{eqn:de_green}
\end{equation}
where $G^{s}(\mathrm{i}\omega_n) = [G_0^{-1}(\mathrm{i}\omega_n) - \bar{\Sigma}^{s}]^{-1}$, while $\im \omega_n$ are fermionic Matsubara frequencies.
Note that, while for each stationary state $s$ the self-energy $\bar\Sigma^s$ is static, the effective DE self-energy $\Sigma_{\text{DE}}(\mathrm{i}\omega_n) = G_0^{-1}(\mathrm{i}\omega_n) - G_{\text{DE}}^{-1}(\mathrm{i}\omega_n)$ is a frequency-dependent object.

\begin{figure*}[t]
   \centering
   \includegraphics[width=\textwidth]{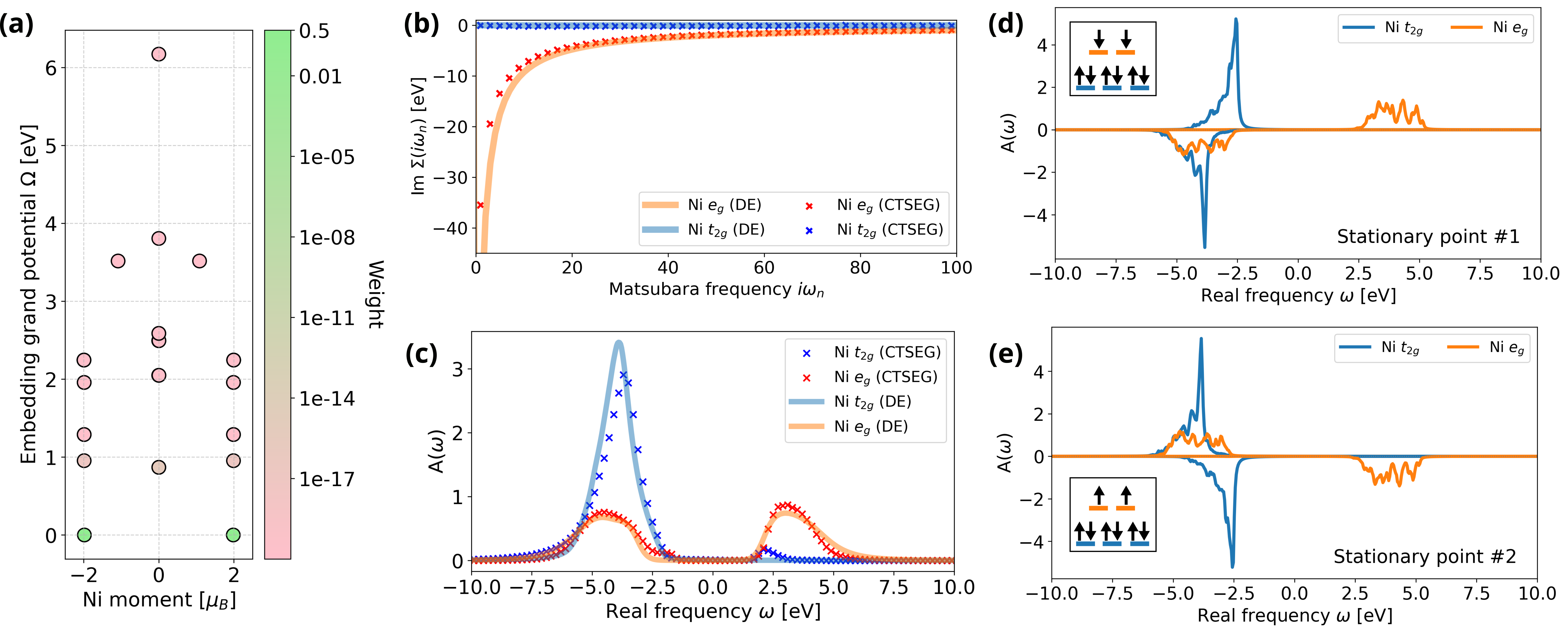}
   \caption{Paramagnetic state of NiO:  (a) embedding grand potential $\Omega$ landscape. Marker color scales with Boltzmann weight; green circles identify the global minima that overwhelmingly dominate the ensemble average for this system; (b) imaginary part of the Matsubara self-energy, $\text{Im}\Sigma(\mathrm{i}\omega_n)$; (c) spin-degenerate local spectral function, $A(\omega)$, obtained via analytic continuation; solid lines represent DE results and crosses denote continuous-time quantum Monte Carlo (CTSEG) benchmarks. Calculations employ a 5-orbital $d$-band model ($U=6$ eV, $J=1$ eV, $\beta=40$ eV$^{-1}$); (d,e) symmetry-broken spectral functions corresponding to the two lowest-energy stationary states from (a) resolving the spin-up (positive y-axis) and spin-down (negative y-axis) components.
   \label{fig:nio_big_comparison}}
\end{figure*}

Importantly, \pref{eqn:de_green} is mathematically analogous to the definition of the average Green's function of a disordered alloy in the coherent-potential approximation (CPA)~\cite{Velicky1968_singlesite_noted,kakehashi_electron_2004}. While single-site CPA averages over an alloy with specific chemical concentrations, the DE acts as a disordered alloy of symmetry-broken static solutions, where the scattering potentials are provided by the static self-energies $\bar{\Sigma}^s$, and their concentrations are given by their thermodynamic weights $p_s = \mathcal{Z}_\text{loc}^{-1} \exp(-\beta \Omega[\bar{\Sigma}^s])$. 

In the particular case of DMFT, the correlated atom is mapped onto a self-consistently determined Anderson impurity model~\cite{georgesDynamicalMeanFieldTheory1996, kotliar_electronic_2006}; the local partition function is then, in this instance, precisely the impurity partition function. This identification provides a direct theoretical link to the polymorphous description of Mott insulators~\cite{trimarchi_polymorphous_2018a, zunger_symmetry_2026}: when brought to self-consistency, DMFT in the DE approximation becomes mathematically identical to a spatially disordered network of local symmetry‑broken HF states treated within single‑site CPA.
A rigorous derivation of \pref{eqn:de_green} for the DMFT case, starting from the Hubbard-Stratonovich decoupling~\cite{altland_condensed_2010a} of the action integral, is given in Appendix~A.

We remark that the DE is not a full multi-determinant treatment: constructing the Green’s function as a statistical average neglects tunneling between basins, and therefore cannot describe the off-diagonal coherences of a true multi-reference state; hence the name ``dephased" ensemble.
This approximation becomes physically sound in the limit of strong Hubbard $U$, where the height of the tunneling barriers separating the basins scales with $U$, suppressing inter‑basin tunneling.
We discuss in more detail the validity of the approximation and strategies for systematic improvement towards the end of the letter.

To evaluate the physical implications of the DE, we investigate two paradigmatic Mott insulators: NiO and FeO. Importantly, in both cases we employ the minimal unit cell, containing a single transition-metal and oxygen atom.
We focus strictly on the paramagnetic regime, comparing one-shot (non charge-self consistent) DFT+DMFT and DFT+DE calculations based on a PBEsol~\cite{perdewRestoringDensityGradientExpansion2008} baseline. Calculations are performed with the Quantum ESPRESSO software distribution~\cite{giannozzi_advanced_2017b, giannozzi_2009_quantumespresso}, while the low-energy tight-binding model is extracted using Wannier90~\cite{marzari_maximally_2012b, pizzi_wannier90_2020c} by including only the 5 bands with majority transition-metal $d$ character close to the Fermi level. For the interaction Hamiltonian we use $U=6$ eV and $J=1$ eV. Computational details on each benchmark are provided in the SI.
While the focus of this paper is the comparison between on-site DE and DMFT, which are both single-site theories, we provide in the SI a direct comparison of the Ni and Fe Green's function computed from the DE ensemble and from an explicit disordered supercell alloy model 
(Sec. IIC and IIIC), to further show the equivalence with the polymorphous description.

Both for ease of implementation and to guarantee the best possible comparison, the DFT+DE calculations are performed in practice by running a HF solver on the DFT+DMFT impurity problem multiple times with different initializations, adapting the protocol of Ponet \textit{et al.}~\cite{ponet_energy_2024} to thoroughly explore the landscape of static solutions (see SI Sec. IC1 for details). For each stationary state $s$, we evaluate $\Omega[\bar{\Sigma}^s]$ to determine its Boltzmann weight and construct the DE Green's function [Eq.~\eqref{eqn:de_green}], which allows us to extract an effective self-energy, $\Sigma_{\text{DE}}(\mathrm{i}\omega_n)$. The self-energy then is embedded back into the crystal and this process is iterated until self-consistency. We benchmark this approximation against DMFT using continuous-time quantum Monte Carlo (QMC)~\cite{kotliar_electronic_2006, georgesDynamicalMeanFieldTheory1996, seth_triqs_2016}, utilizing the segment-picture (CTSEG) solver~\cite{kavokine_ctseg_2025}. Both DMFT and DE calculations are performed using the \texttt{solid\_dmft} package~\cite{merkel_solid_dmft_2022}, built on top of the TRIQS ecosystem~\cite{parcollet_triqs_2015, PhysRevB.105.235115, KAYE2022108458, Kaye2024, Kiese2025}.
We remark that,  to guarantee a faithful numerical comparison with the QMC solvers (and also due to implementation constraints), the actual calculations presented here are, both for DE and QMC, always performed on the Matsubara axis. Nevertheless, DE could in principle allow for a direct evaluation on the real axis.


We first present our results for NiO, characterized by a $d^8$ configuration with a fully occupied $t_{2g}$ shell and a half-filled $e_g$ manifold.
\pref{fig:nio_big_comparison}(a) displays the converged $\Omega[\bar{\Sigma}^s]$ landscape as a function of the local Ni magnetic moment. Out of the many stationary states found, the ensemble statistics is overwhelmingly dominated by the two lowest-energy configurations, which exhibit saturated local moments of approximately $\pm 2\mu_B$.
We remark that we obtain two low-lying states because we impose a spin quantization axis; full rotation invariance would lead to a continuum of degenerate symmetry broken states. The treatment of continuous symmetries is discussed in Sec. ID of the SI.
The next available stationary states lie approximately 1 eV higher in $\Omega$; at room temperature ($\beta = 40$ eV$^{-1}$), their Boltzmann weights are exponentially suppressed by a factor of $\sim 10^{-13}$, rendering their contribution to the Green's function negligible.

In \pref{fig:nio_big_comparison}(b), we compare the DE and DMFT self-energies. Remarkably, the statistical mixture of the symmetry-broken solutions successfully reproduces very closely the features of the DMFT results.
For the $e_g$ bands, the DE successfully captures both the low-frequency Mott pole and the correct high-frequency asymptotics. The completely filled $t_{2g}$ states exhibit a nearly frequency-independent behavior in both DE and the DMFT benchmark.
To obtain the real-frequency lattice spectral function $A(\omega)$ [\pref{fig:nio_big_comparison}(c)], we analytically continue the Matsubara self-energies using the maximum-entropy method~\cite{kraberger_maximum_2017, levyImplementationMaximumEntropy2017}.
The DE and DMFT spectral functions display excellent agreement. Minor quantitative discrepancies are present, such as the slight shift in $t_{2g}$ spectral weight to the conduction band. This is primarily attributable to subtle differences in orbital occupancy: while DE maintains a near-ideal $t_{2g}^6 e_g^2$ configuration, the DMFT results exhibit a fractional transfer of $\sim 0.1$ electrons from the $t_{2g}$ to the $e_g$ shell.
In \pref{fig:nio_big_comparison}(d) and (e), we plot the static, symmetry-broken lattice spectral functions obtained by upfolding the $\bar\Sigma^s$ corresponding to the two lowest-lying $\pm 2\mu_B$ minima, which highlight how for DE, the incoherent Hubbard bands of the paramagnet are captured by a statistical mixture of these static, symmetry-broken insulating states.
We note that a superposition of two HF solutions has been occasionally used in the DMFT literature as an ansatz~\cite{georgesDynamicalMeanFieldTheory1996, kotliar_electronic_2006, rozenberg_motthubbard_1992a}; DE generalizes this heuristic consideration to a weighted mixture of all HF solutions.

\begin{figure}[htbp]
   \centering
   \includegraphics[width=\columnwidth]{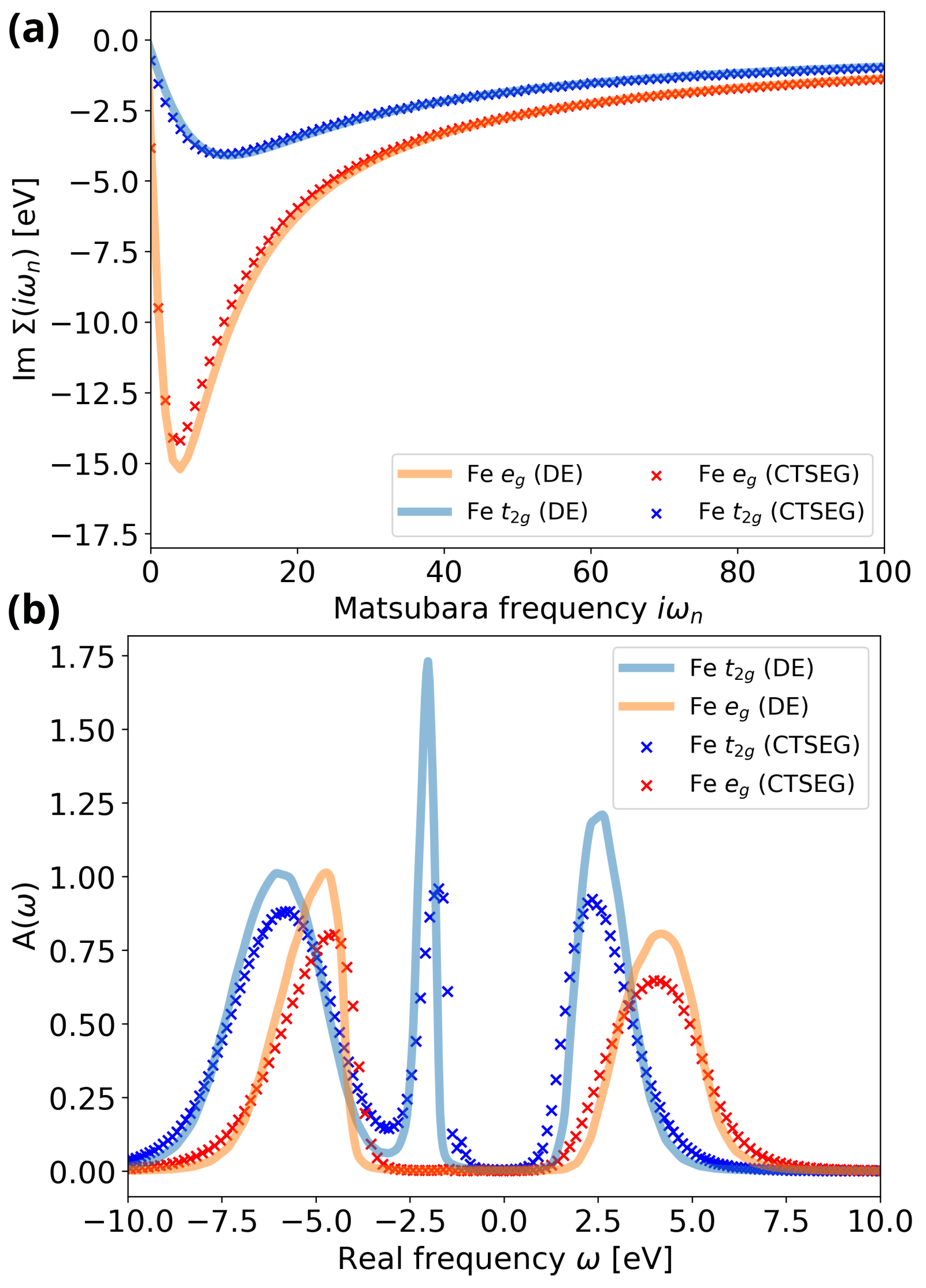}
   \caption{ Paramagnetic state of FeO, 5 orbital model with $U$ = 6 eV and $J$ = 1 eV and $\beta=40$ eV$^{-1}$: (a) imaginary part of the Matsubara self-energy $\text{Im}\Sigma(\mathrm{i}\omega_n)$; (b) local spectral function $A(\omega)$ obtained via analytic continuation of the self-energy. DE results (solid lines) are compared against CTSEG benchmarks (crosses). 
   \label{fig:feo_sigma_spectral_function}}
\end{figure}

We now turn to the more complex $d^6$ Mott insulator FeO, characterized by half-filled $e_g^2$ and fractionally occupied $t_{2g}^4$ Fe $d$ shells. The DE landscape (see Sec. II of the SI) is dominated by six degenerate configurations, which reflect the direction of the majority spin and the permutation of the orbital order of the single minority spin among the degenerate $t_{2g}$ orbitals. Averaging over these static configurations restores both macroscopic paramagnetism and cubic symmetry.
\pref{fig:feo_sigma_spectral_function}(a) demonstrates excellent agreement between the DE and DMFT self-energies in Matsubara frequencies across both $e_g$ and $t_{2g}$ shells. The analytically continued real-frequency lattice spectral functions [\pref{fig:feo_sigma_spectral_function}(b)] also show robust agreement. DMFT exhibits broader features and a small shoulder in the lower $t_{2g}$ Hubbard band which lowers slightly the bandgap (the uncertainty inherent in the analytic continuation process makes precise quantitative comparisons difficult). Ultimately, DE accurately recovers the essential features of the FeO paramagnet.


While CTSEG is efficient for five-orbital systems, it is restricted to density‑density interactions ($W_{ijkl} \simeq W_{iikk}\delta_{ij}\delta_{jk}$), which we adopt in DE for consistency. To demonstrate that DE performs equally well with rotationally invariant interactions, we provide in the SI a benchmark on a two-orbital LuNiO$_3$ model using the CT‑HYB solver~\cite{seth_triqs_2016} including spin‑flip and pair‑hopping terms; DE again yields excellent agreement with DMFT, underscoring its applicability to charge‑ordered systems.

Another defining hallmark of strongly correlated Mott insulators is the persistence of two-particle covariances in (imaginary) time~\cite{georgesDynamicalMeanFieldTheory1996}. In the paramagnetic phase, the existence of a local magnetic moment is signaled by a non-vanishing local spin-spin covariance, $\text{cov}[\hat{S}_z(\tau), \hat{S}_z(0)] = \langle \hat{S}_z(\tau) \hat{S}_z(0) \rangle - \langle \hat{S}_z \rangle^2$.
In a standard static mean-field treatment, the covariance reduces entirely to the bare bubble, which for a gapped system decays exponentially for $\tau \to\beta/2$.
In contrast, DE contains both the bubble term and an additional static contribution from the ensemble (see Appendix B), which prevents the covariance from vanishing. 
\pref{tab:table_covariances} reports the spin-spin correlators evaluated at equal time ($\text{cov}[\hat{S}_z(0), \hat{S}_z(0)] \equiv \text{cov}[\hat{S}_z, \hat{S}_z]$) and at long times ($\text{cov}[\hat{S}_z(\beta/2), \hat{S}_z(0)]$); the full $\tau$ dependence is reported in the SI Sec. IIB and IIIB. Both show very good quantitative agreement (6\% for NiO and 4\% for FeO) with the DMFT results.

\begin{table}[ht]
\centering
\begin{tabular}{lccc}
\hline\hline
  & Method & NiO & FeO \\
\hline
$\text{cov}[\hat{S}_z, \hat{S}_z]$ & DE  & $0.989$ & $3.968$ \\
$\text{cov}[\hat{S}_z, \hat{S}_z]$ & DMFT   & $0.934$ & $3.794$ \\
$\text{cov}[\hat{S}_z(\beta/2), \hat{S}_z(0)]$ & DE   & $0.984$ & $ 3.960$ \\
$\text{cov}[\hat{S}_z(\beta/2), \hat{S}_z(0)]$ & DMFT & $0.928$ & $3.786$ \\
\hline\hline
\end{tabular}
\caption{Local spin-spin covariance of the TM site evaluated at the same $\tau$ and at $\tau = \beta/2$ for the paramagnetic phase of NiO and FeO evaluated both in the DE approximation and DMFT using CTSEG for $\beta = 40\,$eV$^{-1}$.
}
\label{tab:table_covariances}
\end{table}


In summary, we have shown that the DE successfully recovers important defining features of strongly correlated Mott insulators; specifically, the frequency-dependent self-energy and the persistence of local moments, in excellent agreement with DMFT.
The DE approximation reveals that key aspects of 
the complex frequency dependence traditionally associated with ``strong correlations" can be interpreted as the mathematical signature of mapping an alloy/ensemble of local static symmetry-broken states onto a symmetry-restoring single-site description. 
In this view, symmetry breaking does not necessarily diminish the need for strong correlations, as has been suggested~\cite{zunger_symmetry_2026, rivera_identifying_2026}; rather, the statistical coexistence of symmetry‑broken states is itself a fundamental manifestation of correlation physics.

The static nature of the DE marks specific physical boundaries of applicability, and at the same time indicates a clear path for systematic improvements. By freezing the imaginary-time dynamics, the approximation currently neglects intra-basin Gaussian fluctuations, which become relevant for smaller gaps. 
Furthermore, the suppression of inter-basin tunneling precludes the description of coherent quasiparticle dynamics; hence, the DE in its current form cannot describe correlated metals or the behavior near the Mott transition. 
These, in turn, could be captured by simple and explicit frequency-dependent formulations in spectral functional theories~\cite{Vanzini2023, Ferretti2024, Savrasov2004} like the dynamical Hubbard functional or, in a quasiparticle approximation, Koopmans' compliant functionals~\cite{Ferretti2014, caserta_dynamical_2025}.
Last, we mention that both DE and DMFT do not consider structural relaxations, while polymorphic supercells do; key questions to be addressed in future work would be related to the timescales of the electronic fluctuations with respect to the ionic displacements, and to the electronic or ionic mechanisms driving symmetry breaking in the supercells~\cite{zunger_symmetry_2026, trimarchi_polymorphous_2018a, varignon_mott_2019, Georgescu2022, Han2018}.
Nevertheless, DE provides great computational advantage over DMFT for large gap Mott insulators, recovering physics that typically requires high-performance clusters and QMC solvers at a fraction of the cost. Looking ahead, the absence of a fermionic sign problem makes the DE highly attractive for complex insulating $f$-electron systems, and implementing DE in real frequencies instead of the Matsubara axis would bypass the need for analytic continuation.
Moreover, the DE's minimal computational footprint makes it also an ideal tool for pre-conditioning the convergence of expensive DMFT impurity solvers~\cite{makaresz_accelerating_2026, zitko_convergence_2009}, or for scaling into cluster frameworks to directly capture non-local spatial correlations.
In both the case of clusters and $f$-electron systems, the landscape of stationary solutions is expected to show a much more complex structure, with many competing low-energy minima, requiring more complex strategies~\cite{srinivas_gaussian_2010, kleinberg_nearly_2004, chakrabarti_mortal_2008} which we will tackle in upcoming work.

Besides its practical advantages, the DE formalism hints at parallels between the physics of correlated and disordered systems. In Parisi's replica method~\cite{parisi_infinite_1979}, disorder generates an effective interaction term among replicas; conversely here, the quartic interaction term is mapped onto a disordered alloy of symmetry‑broken static solutions. While the nature of the quartic coupling differs, this inverse mapping suggests a deeper duality between correlation effects and self‑generated disorder, and raises the question of whether a more profound connection can be drawn and if static disorder plays a role in other correlated phenomena beyond Mott physics, as hinted in parallel work~\cite{Quinzi2026}.

A.C., M. S. and N.M. acknowledge support by NCCR MARVEL, a National Centre of Competence in Research, funded by the Swiss National Science Foundation (Grant numbers 205602 and 200020\_213082). All calculations presented in this work were performed on the Thor cluster at PSI.
H.U.R.S acknowledges financial support from the Swedish Research Council (Vetenskapsrådet, VR) grant number 2024-04652 and funding from the European Research Council (ERC) under the European Union’s Horizon 2020 research and innovation programme (grant agreement No.\ 854843-FASTCORR).
A.C. and N.M. extend their deep gratitude to Alfredo Fiorentino, Mario Caserta, Libor Vojáček, Edward Linscott, Nicola Colonna, Massimo Capone and especially Matteo Quinzi and Alex Zunger for the insightful conversations.

\clearpage

\section*{End Matter}

\subsection{Functional Integral Derivation of DMFT in the DE approximation}

To rigorously derive \pref{eqn:de_green} for the specific case of DMFT, we consider a correlated atom in the interacting subspace which is mapped onto a self-consistent Anderson impurity model, where the many-body physics of the local embedding problem physics is captured by the effective Hamiltonian $\hat{H}_{\text{imp}}$~\cite{georgesDynamicalMeanFieldTheory1996, kotliar_electronic_2006, held_electronic_2007a}:

\begin{align}
    &\hat{H}_{\text{imp}} = \sum_{ij, \sigma} [\epsilon_{ij} - (\mu+V_\text{DC})] \hat{\psi}_{i\sigma}^\dagger \hat{\psi}_{j\sigma} + \sum_{i \kappa, \sigma} (V_{i\kappa} \hat{\psi}_{i\sigma}^\dagger \hat{b}_{\kappa \sigma} + \text{h.c.}) \label{eqn:h_imp} \\
    &+ \sum_{\kappa, \sigma} \varepsilon_{\kappa} \hat{b}_{k\sigma}^\dagger \hat{b}_{\kappa \sigma} + \frac{1}{2} \sum_{ijkl, \sigma\sigma'} W_{ijkl} \hat{\psi}_{i\sigma}^\dagger \hat{\psi}_{k\sigma'}^\dagger \hat{\psi}_{l\sigma'} \hat{\psi}_{j\sigma} . \nonumber
\end{align}

The first term represents the local one-body part, where $\epsilon_{ij}$ denotes the on-site energy levels and $\mu$ is the chemical potential and $V_\text{DC}$ is the double counting potential, expressed as a derivative of the double counting energy $E_\text{DC}$, i.e. $V_{DC}= \delta E_{DC}/\delta N$; the $\psi_{i\sigma}, \psi^\dagger_{i\sigma}$ represent fermionic Grassmann fields for each orbital index $i$ and spin index $\sigma$. The second sum characterizes the hybridization with the bath through matrix elements $V_{i\kappa}$. The interaction between electrons is described by the static four-point tensor $W_{ijkl}$, while $E_{DC}$ is the double-counting correction energy, which in this work is always taken to be the fully-localized limit~\cite{cococcioni_linear_2005, ryee_effect_2018}. From here on, we absorb the orbital and spin index into a compound index $(i, \sigma) \rightarrow \mathbf{i}$.
The partition function $\mathcal{Z}_{\text{imp}}$ 
can be evaluated by performing a Hubbard-Stratonovich (HS) transformation $\exp[-\frac{1}{2}\psi^*\psi\,\mathcal{U}\,\psi^*\psi] \propto 
\int\!\mathcal{D}\varphi\, \exp[-\frac{1}{2}\varphi\mathcal{U}^{-1}\varphi - \mathrm{i}\varphi\psi^*\psi]$ introducing an auxiliary bosonic field $\varphi_{m\mathbf{ij}} \equiv \varphi_{\mathbf{ij}}(\mathrm{i}\nu_m)$ defined on the bosonic Matsubara frequencies $\im \nu_m$ and the antisymmetrized interaction tensor $\mathcal{U}_{\mathbf{ijkl}} = W_{\mathbf{ijkl}} - W_{\mathbf{ilkj}}$. The term $\mathcal{A}$ is the Hubbard-Stratonovich action and reads:
\begin{align}
    \label{eqn:hs_action_split} 
    \mathcal{A} &= \underbrace{\sum_{n, \mathbf{ij}} \psi^*_{n\mathbf{i}} [G_0]^{-1}_{\mathbf{ij}}(\mathrm{i}\omega_n) \psi_{n\mathbf{j}}  
    + \mathrm{i} \sum_{n,m, \mathbf{ij}} \varphi_{m\mathbf{ij}} \psi^*_{n\mathbf{i}} \psi_{n+m,\mathbf{j}} \nonumber}_{\mathcal{A}_F}\\  &+
    \underbrace{\frac{1}{2} \sum_{m, \mathbf{ijkl}} \varphi_{m\mathbf{ij}} \mathcal{U}^{-1}_{\mathbf{jilk}} \varphi_{-m,\mathbf{kl}}}_{\mathcal{A}_B}
\end{align}
where $\psi_{\mathbf{i}}(\im \omega_n)$ are the Fourier components of the fermionic Grassmann fields. In DMFT the non-interacting part $G_0$ is termed ``Weiss field"~\cite{georgesDynamicalMeanFieldTheory1996, kotliar_electronic_2006} and can be written as $(G_0)_{\mathbf{ij}}(\mathrm{i}\omega_n) = [\mathrm{i}\omega_n \delta_{\mathbf{ij}} - \epsilon_{\mathbf{ij}} + V_{DC}\delta_\mathbf{ij} + \mu \delta_\mathbf{ij} - \Delta_{\mathbf{ij}}(\mathrm{i}\omega_n)]^{-1}$, where $\Delta_{\mathbf{ij}}$ is the bath hybridization function.
The single-particle Green's function of the impurity problem can be generally defined as the functional integral over Grassmann fields~\cite{altland_condensed_2010a}:
\begin{equation}
    G(\mathrm{i}\omega_n) = \frac{1}{\mathcal{Z}_{\text{imp}}} \int \mathcal{D}[\psi^*, \psi,\, \varphi] \, \psi_{n} \psi^*_{n} \, e^{-\mathcal{A}[\psi^*, \psi, \, \varphi]} \, .
\end{equation}
A standard saddle‑point decoupling of the impurity action consists in restricting the auxiliary field to static configurations ($\varphi(\tau) \to \bar{\varphi}$) and approximating the functional integral by a single field configuration that makes the action stationary, $\delta \mathcal{A} / \delta \bar{\varphi} = 0$.
This is mathematically equivalent to the HF decoupling of the action~\cite{altland_condensed_2010a}; when iterated to self‑consistency within the DFT+DMFT cycle it therefore becomes identical to DFT+$U$~\cite{kotliar_electronic_2006, carta_explicit_2025}.
The DE approximation precisely consists in relaxing the single‑saddle‑point restriction and instead approximates the functional integral over $\mathcal{D}[\bar{\varphi}]$ as a discrete sum over all static fields $\bar{\varphi}^s$ that satisfy the stationary condition $\delta \mathcal{A} / \delta \bar{\varphi} = 0$.
For a fixed configuration $\bar{\varphi}^s$, the purely bosonic part of the action $\mathcal{A}_B = \frac{\beta}{2} \mathrm{Tr}[\bar{\varphi}^s \mathcal{U}^{-1} \bar{\varphi}^s]$, factors out of the Grassmann integrals, leading to:
\begin{equation}
    \label{eqn:definition_of_Z_DE}
    Z_\text{imp} \simeq \sum_s e^{-\mathcal{A}_B[\bar\varphi^s]} \int \mathcal{D}[\psi^*, \psi] \,  \, e^{-\mathcal{A}_F[\psi^*, \psi, \, \bar\varphi^s]} \,  , \\
\end{equation}
and
\begin{equation}
    \label{eqn:definition_of_G_DE}
    G_\text{DE}(\mathrm{i}\omega_n) =\sum_s \frac{e^{-\mathcal{A}_B[\bar\varphi^s]}}{\mathcal{Z}_{\text{imp}}} \int \mathcal{D}[\psi^*, \psi] \, \psi_{n} \psi^*_{n} \, e^{-\mathcal{A}_F[\psi^*, \psi, \, \bar\varphi^s]} \, .
\end{equation}
The remaining fermionic part $\mathcal{A}_{F}$ evaluates to the fermion determinant, leading to the integrals~\cite{altland_condensed_2010a}:
\begin{align}
    &\int \mathcal{D}[\psi^*, \psi] e^{-\mathcal{A}_F} = \det \left[ (G^s)^{-1} \right] = e^{-\text{Tr} \ln [ G^s ]} \label{eqn:fermion_det}\\
    &\int \mathcal{D}[\psi^*, \psi] \, \psi_{n} \psi^*_{n} \, e^{-\mathcal{A}_F} = G^s(\im \omega_n) \, e^{-\text{Tr} \ln [ G^s]} \label{eqn:fermion_gf}
\end{align}
where $(G^s)^{-1} = G_0^{-1} + \mathrm{i}\bar{\varphi}^s$, here one can see that the field $\bar\varphi^s$ takes the form of a self-energy term $\bar\Sigma^s = -\im \bar\varphi^s$.
Combining \pref{eqn:definition_of_Z_DE} with \pref{eqn:fermion_det} and making the bosonic action $\mathcal{A}_B$ explicit, the total statistical weight for a given configuration can be written as $\exp(-\beta \Omega[\bar{\varphi}^s])$, defining the embedding grand potential:
\begin{equation}
    \Omega[\bar{\varphi}^s] = \frac{1}{\beta} \text{Tr} \ln \left[ G^s \right] + \frac{1}{2} \mathrm{Tr} [ \bar{\varphi}^s \mathcal{U}^{-1} \bar{\varphi}^s] \,  .
\end{equation}
Using the stationarity condition $\delta\mathcal{A}/\delta\bar\varphi=\delta\Omega/\delta\bar\varphi=0$ recovers the HF self-energy $\bar\Sigma^s= \mathcal{U} \bar{\rho}^s$. The quadratic bosonic term then reduces to minus the Hartree-Fock interaction energy $-\frac{1}{2} \mathrm{Tr}[\bar{\rho}^s \mathcal{U} \bar{\rho}^s]$, recovering \pref{eqn:grand_pot_hf}. Finally, substituting \pref{eqn:fermion_gf} into the ensemble sum yields the DE Green's function presented in \pref{eqn:de_green}.
We note that, while here we have presented the derivation of the DE approximation on top of the DMFT impurity problem, our approach could also be derived from a particular flavor of CPA described by Kakehashi \textit{et al.}~\cite{kakehashi_electron_2004, kakehashi_firstprinciples_2010, kakehashi_manybody_2002} by restricting the thermodynamic integration to the paths of stationary action.

\subsection{Evaluating two-point covariances}
\label{sec:appendix_covariances}

To characterize the fluctuations within the ensemble, we consider two general single-particle operators $\hat{\mathcal{O}}(\tau) = \sum_{\mathbf{ij}} \mathcal{O}_{\mathbf{ij}} \hat{c}^\dagger_{\mathbf{i}}(\tau) \hat{c}_{\mathbf{j}}(\tau)$ and $\hat{\mathcal{Q}}(\tau) = \sum_{\mathbf{kl}} \mathcal{Q}_{\mathbf{kl}} \hat{c}^\dagger_{\mathbf{k}}(\tau) \hat{c}_{\mathbf{l}}(\tau)$. We evaluate the two-point covariance or ``connected correlator": 
\begin{equation}
\text{cov}[\hat{\mathcal{O}}(\tau), \hat{\mathcal{Q}}(0)] = \langle \hat{\mathcal{O}}(\tau) \hat{\mathcal{Q}}(0) \rangle - \langle \hat{\mathcal{O}} \rangle \langle \hat{\mathcal{Q}} \rangle \, ,
\end{equation}
where in the DE approximation the average is $\langle \cdot \rangle = \sum_s p_s \langle \cdot \rangle_s$ with weights $p_s = \mathcal{Z}_{\text{imp}}^{-1} \exp(-\beta \Omega[\bar{\varphi}^s])$.
Since each stationary point $s$ corresponds to a quadratic action, we apply Wick's theorem to the four-operator expectation value within each configuration:
\begin{align}
    \langle \hat{\mathcal{O}}(\tau) \hat{\mathcal{Q}}(0) \rangle_s &= \sum_{\mathbf{ijkl}} \mathcal{O}_{\mathbf{ij}} \mathcal{Q}_{\mathbf{kl}} \langle \hat{c}^\dagger_{\mathbf{i}}(\tau) \hat{c}_{\mathbf{j}}(\tau) \hat{c}^\dagger_{\mathbf{k}}(0) \hat{c}_{\mathbf{l}}(0) \rangle_s \nonumber \\
    &= \langle \hat{\mathcal{O}} \rangle_s \langle \hat{\mathcal{Q}} \rangle_s - \sum_{\mathbf{ijkl}} \mathcal{O}_{\mathbf{ij}} \mathcal{Q}_{\mathbf{kl}} G^s_{\mathbf{jk}}(\tau) G^s_{\mathbf{li}}(-\tau) \, ,
\end{align}
where the first term is the direct contraction and the second is the bare susceptibility (bubble) contribution $\chi^s_{\mathbf{jilk}}(\tau) = G^s_{\mathbf{jk}}(\tau) G^s_{\mathbf{li}}(-\tau)$. Substituting this into the definition of the covariance, we obtain:
\begin{align}
    \text{cov}[\hat{\mathcal{O}}(\tau), \hat{\mathcal{Q}}(0)] &= \sum_s p_s \left( -\sum_{\mathbf{ijkl}} \mathcal{O}_{\mathbf{ij}} \chi^s_{\mathbf{jilk}}(\tau) \mathcal{Q}_{\mathbf{kl}} \right) \nonumber \\
    &+ \left[ \sum_s p_s \langle \hat{\mathcal{O}} \rangle_s \langle \hat{\mathcal{Q}} \rangle_s - \langle \hat{\mathcal{O}} \rangle \langle \hat{\mathcal{Q}} \rangle \right] \, . \label{eqn:de_cov_split}
\end{align}
The first line of Eq. \eqref{eqn:de_cov_split} represents the ensemble-averaged dynamical bubble, describing fluctuations within individual basins. The second line, in brackets, is the ensemble variance. This term captures the static (i.e. $\tau-$independent) correlations arising from the variation of mean-field observables across the potential landscape. 
In the single-saddle point limit $s = \circ$, the ensemble variance identically vanishes, leaving only the bare bubble $\chi^\circ$. In the presence of a band gap, this bubble term decays exponentially at large imaginary times $\tau \to \beta/2$. 
Therefore, in an insulating state described at the static mean-field (HF) level, all observable two-particle correlations are lost for large enough imaginary time~\cite{georgesDynamicalMeanFieldTheory1996}. In this specific sense, HF cannot describe correlations.
Conversely, in the ensemble picture, while the bubble term still decays to zero at large $\tau$, the static ensemble variance remains as the dominant, non-vanishing contribution and correlations persist at all $\tau$.

\bibliography{alberto_bibliography, other_references}
\end{document}